\documentclass[%
 reprint,
superscriptaddress,
 amsmath,amssymb,
 aps,
]{revtex4-2}
\usepackage[T1]{fontenc}
\usepackage[utf8]{inputenc}
\usepackage{graphicx}
\usepackage{dcolumn}
\usepackage{bm}
\usepackage{braket}
\usepackage{xcolor}
\usepackage{pdfpages}
\usepackage{pgffor}
\makeatletter
\AtBeginDocument{\let\LS@rot\@undefined}
\makeatother
\usepackage[colorlinks=true, urlcolor=blue, linkcolor=blue, citecolor=blue]{hyperref}

\newcommand{\oratomic}{Oratomic, Pasadena, California, 91125}
\newcommand{\berkeleyphy}{Department of Physics, University of California, Berkeley, California 94720}
\newcommand{\CIQC}{Challenge Institute for Quantum Computation, University of California, Berkeley, California 94720}
\newcommand{\LBL}{Materials Sciences Division, Lawrence Berkeley National Laboratory, Berkeley, California 94720}
\newcommand{\columbia}{Department of Physics, Columbia University, New York, New York 10027}
\begin{document}

\preprint{APS/123-QED}

\title{A Rydberg atom array coupled to a high-cooperativity optical cavity}
\author{Tai Xiang}
\affiliation{\berkeleyphy}
\affiliation{\CIQC}

\author{Yue-Hui Lu}
\affiliation{\berkeleyphy}
\affiliation{\CIQC}

\author{Tsai-Chen Lee}
\affiliation{\berkeleyphy}
\affiliation{\CIQC}

\author{Jacquelyn Ho}
\affiliation{\berkeleyphy}
\affiliation{\CIQC}

\author{Michael Lan}
\affiliation{\berkeleyphy}
\affiliation{\CIQC}

\author{Zhenjie Yan}
\affiliation{\berkeleyphy}
\affiliation{\columbia}

\author{Harry Levine}
\affiliation{\berkeleyphy}
\affiliation{\CIQC}
\affiliation{\oratomic}

\author{Nathaniel B. Vilas}
\affiliation{\berkeleyphy}
\affiliation{\CIQC}

\author{Dan M. Stamper-Kurn}
\email[]{dmsk@berkeley.edu}
\affiliation{\berkeleyphy}
\affiliation{\CIQC}
\affiliation{\LBL}

\begin{abstract}
Integrating Rydberg-state control of atom-atom interactions with cavity control of atom-light interactions enables the combination of local entangling operations with photon-mediated long-range coupling, measurement, and communication. Such integration is challenging because electric fields generated by charges on cavity optics and by cavity-stabilizing actuators may disturb Rydberg-state coherence. Here, we couple a Rydberg tweezer array to a
high-cooperativity optical cavity, suppressing the
electric fields produced by cavity length actuation so that the residual field variation of about $3~\mathrm{mV/cm}$ rms does not
limit Rydberg coherence. We demonstrate two capabilities that neither
system provides alone. First, we use Purcell-enhanced fluorescence to read out atoms
directly in the ground--Rydberg basis within $7.5~\mu\mathrm{s}$, well within
the Rydberg-state lifetime, with ground- and Rydberg-state detection
infidelities of $3\%$ and $5\%$, respectively, and with low atom loss. Second, we use Rydberg-Rydberg interactions to control collective light--atom
scattering into the cavity mode.  Specifically, we drive blockaded Rabi
oscillations between a state with two ground-state atoms, which scatter collectively with a superradiant or subradiant rate controlled by the atomic positions, and an entangled state with a single Rydberg excitation, which scatters like a single atom. These capabilities open
routes toward networked Rydberg processors, mid-circuit measurements in analog quantum simulators, and many-body quantum optics with entangled emitters coupled collectively to light.
\end{abstract}

\maketitle
Control of atom–atom interactions in neutral-atom systems has advanced rapidly through the use of Rydberg states to couple atoms held in optical tweezer arrays \cite{Saffman2010_rev, Kaufman2021OpticalTweezers}, enabling applications in quantum computing \cite{Bluvstein2024logical,saffman2026neutralatomquantumcomputing, Henriet_2020, graham2022multiqubit}, many-body quantum simulation and state preparation~\cite{Bernien2017ManyBody,scholl2021quantum,browaeys2020many, Omran2019_cat}, quantum optimization \cite{Ebadi2022QuantumOptimization, Oliveira2025graph}, and sensing \cite{Adams2020Rydberg,Holloway2014, eckner2023realizing, bornet2023scalable}. In parallel, optical cavities provide a powerful means of controlling atom–photon interactions by enhancing the coupling of atoms to selected modes of light. Optical cavities have enabled quantum-enhanced metrology \cite{MSS2010_cavityfeed,cox2016_squeeze, hosten2016measurement, pedrozo2020entanglement}, rapid nondestructive and mid-circuit measurement \cite{Volz2011Measurement, DeistMCM, mcmvladan, Lee2026Rapid}, coherent atom–photon interfaces \cite{Reiserer2015CavityQuantumNetworks,kimble2008quantum}, and photon-mediated long-range interactions between atoms \cite{Mivehvar02012021}.

Theoretical proposals have explored combining Rydberg-mediated atom–atom interactions with strong atom-photon coupling in optical cavities. Such systems could realize Rydberg quantum processors networked through photonic links \cite{young2022architecture,covey_quantum_2023, sinclair2025link,li2024interconnect} and many-body dynamics in which finite-range Rydberg interactions compete with collective cavity-mediated interactions \cite{mendonca25,liang2026_DI,hosseinabadi2026_RydQED,puel24}.

However, optical cavities present a challenging setting for controlling Rydberg atoms with large principal quantum number $n$.  Strong coupling is typically achieved using small-volume resonators, in which nearby optical elements and cavity-stabilization actuators may impose stray electric fields on and perturb Rydberg atoms \cite{ocola_control_2024,raj_surface_2026}, whose extreme polarizability scales as $n^7$. 
Techniques to mitigate these fields have not previously been achieved in systems that reach the single-atom strong coupling regime \cite{ocola_control_2024, ningyuan_observation_2016,DeSantis2026CavityRydberg,sheng_intracavity_2017}.

Here, we report the integration of a Rydberg tweezer array with a strongly coupled optical cavity. We characterize and control electric fields associated with the cavity, achieving coherent excitation of individual atoms and Rydberg-blockaded atom pairs within the cavity. We leverage the system for two capabilities central to quantum control.  First, we use Purcell-enhanced cavity fluorescence to perform fast, nondestructive detection of atoms in ground and Rydberg states, providing a path toward mid-circuit measurement and feedback control directly in the ground–Rydberg basis. Second, we demonstrate the simultaneous control of Rydberg interactions and collective atom–light coupling. By varying the positions of atoms with respect to each other and to the cavity mode, we realize blockaded and unblockaded Rydberg dynamics together with superradiant and subradiant optical responses. This independent control enables the study of many-body dynamics arising from the interplay of local and global interactions.

\begin{figure}[t]
    \centering
    \includegraphics[width=1\linewidth]{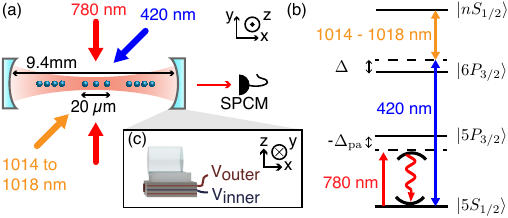}
    \caption{Experimental platform. (a) A one-dimensional array of $^{87}\mathrm{Rb}$ atoms is positioned at the center and along the axis of a 1-cm-long near-concentric Fabry–P\'{e}rot cavity. The atoms are addressed by Rydberg-excitation and probe beams oriented at $45^\circ$ to one another and with wavelengths as shown. Cavity emission is detected with a single-photon counting module (SPCM). Up to three atoms at the cavity center are uniformly excited to Rydberg states, while an additional supply of atoms outside the excitation region is used to replenish depleted sites. (b)  Atoms are driven on a two-photon transition from the $|5S_{1/2}, F=2, m_F = 2\rangle$ atomic ground state to $|n S_{1/2}, m_J=1/2\rangle$ Rydberg states ($n$ between 41 and 60)  via the $|6P_{3/2}, F=3 \rangle$ intermediate state, with an intermediate-state detuning of $\Delta \approx 2 \pi \times 780$ MHz.  Retro-reflected probe light is detuned by $-\Delta_\mathrm{pa}$ from the $|5S_{1/2}, F=2\rangle \rightarrow |5P_{3/2}, F'=3\rangle$ $D_2$ atomic transition, and resonant with a TEM$_{00}$ mode of the optical cavity. (c) Illustration of the cavity mirror, mounting structure, and the shear piezoelectric actuators used for cavity-length stabilization and electric-field compensation. Independent actuator control voltages are labeled.}
    \label{fig:scheme}
\end{figure}

Our experiments use a one-dimensional chain of $^{87}\mathrm{Rb}$ atoms confined in optical tweezers $808$~nm in wavelength and coupled to a near-concentric Fabry–P\'erot cavity mode~\cite{yan2023superradiant}. The system operates in the single-atom strong-coupling regime on the $^{87}\mathrm{Rb}$ $D_2$ cycling transition, with cooperativity $C = g_0^2/(2\kappa \gamma) = 3.0$, where $2g_0$ is the single-photon Rabi frequency for the $|F=2, m_F=\pm 2\rangle \rightarrow |F'=3,m_F'=\pm 3\rangle$ cycling transition, and $\kappa$ and $\gamma$ are the cavity and atomic half-linewidths, respectively. The corresponding parameters are $\{g_0,\kappa,\gamma\}=2\pi \times \{3.1, 0.53, 3.0\}$~MHz. We drive two-photon excitations to Rydberg states ranging from $n=41$ to $n=60$ using two counter-propagating optical fields incident at an angle of $45^\circ$ relative to the cavity ($x$) axis (Fig.\ \ref{fig:scheme}(a-b)). We switch off the tweezers during Rydberg excitation to avoid Rydberg anti-trapping and tweezer-induced ac Stark shifts.  Ground-state atoms are detected by fluorescence on the $D_2$ transition at 780~nm wavelength, driven by counter-propagating beams perpendicular to the $x$ axis.  Fluorescence is detected either by free-space imaging or, as described below, by the emission of cavity photons.  For experiments using free-space imaging, we use a zoned architecture, separating atoms between storage and Rydberg-excitation zones, enabling repeated measurements within a single loading cycle.  Further details are provided in the Supplemental Material \cite{supplemental}.

We first characterize and control the static electric fields within our optical cavity. 
When probing the single-atom Rydberg resonance without electric field suppression measures, we detect strong fields that emerge from two sources: surface charging of nearby (as close as $4.7$ mm) insulating surfaces, and stray fields emitted by piezoelectric actuators used to stabilize the cavity length; see Fig.\ \ref{fig:scheme}(c).  

We mitigate surface charging using ultraviolet illumination \cite{ziemba_removal_2025,ocola_control_2024,mamat2024_charge} (see Supplemental Material for details~\cite{supplemental}). The remaining field originates dominantly from the unshielded piezoelectric actuators in our setup, whose time-dependent drive voltages generate electric fields at the atoms along ($x$) and transverse to ($z$) the cavity axis, thereby coupling cavity length changes directly to the Rydberg resonance frequency.  

\begin{figure}[t]
    \centering
    \includegraphics[width=1\linewidth]{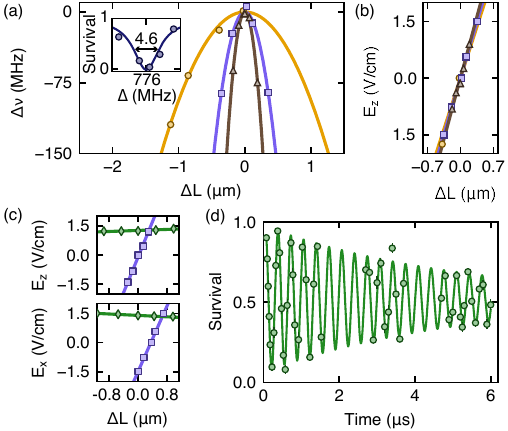}
    \caption{Electric-field compensation and characterization. (a) Rydberg transition frequency shift, $\Delta \nu$, versus cavity-length displacement for $n=41$ (circles), $53$ (squares), and $60$ (triangles), with parabolic fits (solid curves). Inset: representative Rydberg excitation spectrum with a Lorentzian fit. (b) Corresponding electric fields inferred from $\Delta \nu$ and the Rydberg state polarizabilities, with linear fits (solid lines). Symbols are the same as those in panel (a). The inferred electric field agrees across all three Rydberg states. (c) Electric fields along the $x$ and $z$ axes versus cavity-length displacement without (purple squares) and with (green diamonds) compensation. Solid lines are linear fits. (d) Coherent $n=53$ Rydberg Rabi oscillations with active cavity-length stabilization and electric-field compensation. The data are averaged over arrays of three atoms separated by $9.4~\mu\mathrm{m}$ to suppress Rydberg interactions. An exponentially decaying sinusoidal fit (solid line) yields a $1/e$ decay time of $5.1(2)~\mu\mathrm{s}$. Error bars denote $68\%$ confidence intervals.}
    \label{fig:Efieldcontrol}
\end{figure}

Indeed, without mitigation, tuning the cavity length $L$ over one free spectral range generates electric fields up to $1.5$ V/cm, producing Stark shifts of hundreds of MHz for $n>53$ (Fig.\ \ref{fig:Efieldcontrol}(a,b)). To suppress these fields, we apply voltages differentially to four actuator electrodes -- the inner and outer electrodes of the two shear piezoelectric actuators (Fig. \ref{fig:scheme}(c)) -- so as to vary the cavity length without producing electric fields along the $x$ and $z$ axes. 
Using the $n=53$ Rydberg state to benchmark the compensation, we find that the dependence of the electric field on cavity length drops by a factor of $50\text{--}70$, showing substantial suppression of the actuator fields; see Fig.\ \ref{fig:Efieldcontrol}(c).
Additionally, comparison of the measured Rydberg resonance frequency with theoretical predictions for several principal quantum numbers reveals no systematic frequency shifts attributable to a residual field along the uncontrolled $y$ axis. Further details of the compensation and field characterization are provided in the Supplemental Material~\cite{supplemental}. Having suppressed the electric fields in our cavity, we observe coherent ground-to-Rydberg Rabi oscillations at a Rabi frequency of $2\pi\times3.0~\mathrm{MHz}$ and a $1/e$ decay time of $5.1(2)~\mu\mathrm{s}$ (Fig.\ \ref{fig:Efieldcontrol}(d)).  Here, Rydberg population is detected by imaging atom loss, with loss mapped to Rydberg excitation through anti-trapping of the Rydberg state by the optical tweezer. 

\begin{figure}[t]
    \centering
    \includegraphics[width=\linewidth]{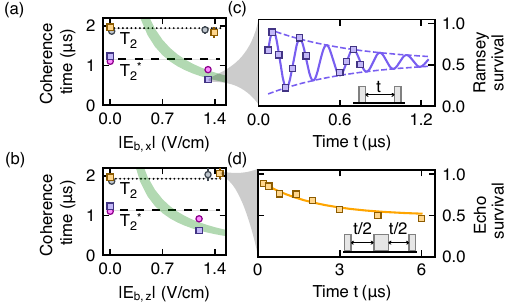}
    \caption{Electric field noise and Rydberg coherence. (a,b) Ramsey ($T_2^*$) and spin-echo ($T_2$) coherence times versus static electric field bias along the $x$ (a) and $z$ (b) axes, measured with (squares) and without (circles) the cavity actively stabilized. Green bands indicate the simulated Ramsey coherence time limited by only the measured electric-field noise $\sigma_E$ under active cavity stabilization (see text). The shaded region reflects the uncertainty propagated from $\sigma_E$. Dashed and dotted lines are guides to the eye. Ramsey data taken without active stabilization are fitted using a maximum-likelihood model accounting for slow electric field drifts (see Supplemental Material~\cite{supplemental}). (c) Representative Ramsey measurement, with a fit to a Gaussian-decaying sinusoid (solid curve). Dashed curves show the decay envelope. (d) Representative spin-echo measurement with an exponential fit. Error bars denote $68\%$ confidence intervals.}
    \label{fig:coherences}
\end{figure}

We characterize the ground-Rydberg coherence more quantitatively through single-atom Ramsey and spin-echo measurements at zero field and with a large bias field $E_b\sim1.3~\mathrm{V/cm}$ applied along either the $x$ or $z$ axis (Fig.~\ref{fig:coherences}). We compare the resulting $T_2^*$ and $T_2$ coherence times with the cavity actively stabilized and with the actuator electrodes held at constant voltage. The applied bias increases the first-order sensitivity of the Rydberg transition to electric-field fluctuations along the bias direction. At zero bias, we measure $T_2^*=1.2(1)~\mu\mathrm{s}$ and $T_2=2.0(1)~\mu\mathrm{s}$, with no discernible dependence on the cavity-stabilization condition. Applying a bias field reduces $T_2^*$, while $T_2$ remains unchanged, indicating electric-field fluctuations that are slow compared with the $\mu\mathrm{s}$-scale experimental sequence. These fluctuations are systematically larger when the cavity is actively stabilized, consistent with imperfect cancellation of the stray electric fields produced by the driven piezoelectric actuators.  

We quantify the residual electric-field noise by modeling Ramsey decay as arising from independent shot-to-shot Gaussian frequency fluctuations. Electric-field fluctuations with rms magnitude $\sigma_E$ limit the coherence time to $T_{2,E}^*=(\sqrt{2}\pi\alpha E_b\sigma_E)^{-1}$, where $\alpha$ is the polarizability expressed in frequency units, such that the quadratic Stark shift is $\Delta\nu=-\frac{1}{2}\alpha E^2$.  Other fluctuations that are independent of $E_b$, such as phase noise of the driving light beams, would themselves limit the coherence time to $T_{2, \mathrm{bg}}^*$. 
Altogether, the observed coherence time at the field $E_b$ is then
\begin{equation}
    T_2^*(E_b) = \left[ \left(\frac{1}{T_{2,\text{bg}}^*} \right)^2 + \left(\frac{1}{T_{2,\text{E}}^*(E_b)} \right)^2 \right]^{-1/2}.
\end{equation}

With this expression, accounting for coherence times observed with and without an applied bias field, we quantify the rms electric field fluctuations along the $x$ and $z$ axes as 
$\sigma_{E_x}=3.5(5)~\mathrm{mV/cm}$ and $\sigma_{E_z}=3.0(4)~\mathrm{mV/cm}$ under active cavity stabilization, compared with $1.5(4)~\mathrm{mV/cm}$ along both axes with the piezoelectric actuators grounded.  Importantly, the inferred electric-field-limited coherence time at zero applied bias field exceeds the limits imposed by Rydberg-state lifetime and Doppler broadening near the zero-field operating point (green curves in Fig.~\ref{fig:coherences}(a, b)). The measured coherence in our system is instead consistent with being dominated by Rydberg-laser phase noise, which can be improved in future work. 

We now leverage strong optical coupling to access capabilities unique to the integrated platform. One such capability is fast, Purcell-enhanced cavity detection of the atomic state. In ground-Rydberg qubit encodings, such as those applied for analog quantum simulation and numerical optimizers and solvers~\cite{Ebadi2022QuantumOptimization,scholl2021quantum}, Rydberg population is typically detected destructively through anti-trapping-induced atom loss. Here, instead, we use cavity-enhanced fluorescence to perform nondestructive state readout within $7.5~\mu$s, shorter than the Rydberg-state lifetime, which could enable measurement and feed-forward directly in the ground–Rydberg basis.

\begin{figure}[t]
    \centering
    \includegraphics[width=1\linewidth]{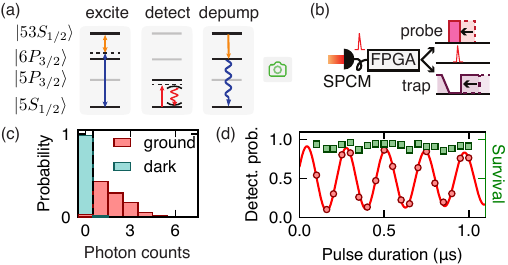}
    \caption{Nondestructive detection of a single Rydberg atom. (a) State-selective Purcell-enhanced fluorescence protocol. After Rydberg excitation, a ground-state atom scatters $780$~nm probe photons into the cavity, while a Rydberg atom remains dark. Following the detection pulse, the remaining Rydberg population is de-excited with a pulse of resonant 1015-nm-wavelength light. The atom is then imaged with fluorescence detection. (b) Adaptive detection scheme. Detection of the first cavity photon terminates the probe and restores the tweezer; otherwise, detection continues for a maximum of $7.5~\mu\mathrm{s}$. (c) Histograms of collected photon counts for a ground-state atom and for an empty tweezer. Dashed line denotes the detection threshold for ground-state atoms. (d) Rydberg Rabi oscillations measured by cavity-based detection (red circles) and corresponding atom survival (green squares). Photon count data are post-selected on atom survival.
    }
    \label{fig:CavDetect}
\end{figure}

Our detection protocol, summarized in Fig.\ \ref{fig:CavDetect}(a), distinguishes ground and Rydberg states through state-selective, cavity-enhanced fluorescence. Following Rydberg excitation, a retro-reflected probe beam $780$~nm in wavelength drives atoms in the $\ket{5S_{1/2}}$ manifold to scatter photons into the cavity, while Rydberg atoms remain dark. Detection of a cavity photon identifies the ground state, whereas no detected photons within the detection window identifies the Rydberg state. After detection, we apply a $2.5~\mu$s-long pulse of resonant light 1015~nm in wavelength to depump any remaining Rydberg population to the ground state before restoring the tweezers. Additional details on the state-detection protocol are included in the Supplemental Material~\cite{supplemental}.

To minimize measurement-induced loss, we adaptively terminate detection once the atomic state is identified~\cite{Lee2026Rapid, mcmvladan, zeng2026fast}. Upon detecting the first cavity photon, a field-programmable gate array (FPGA) switches off the probe and immediately restores the tweezer, limiting probe scattering and free expansion; see Fig.\ \ref{fig:CavDetect}(b). Otherwise, detection continues for $7.5~\mu\mathrm{s}$, followed by the depump pulse and tweezer turn-on. A final fluorescence image independently measures atom survival. The corresponding photon-count distributions are shown in Fig.\ \ref{fig:CavDetect}(c). We measure a ground-state detection infidelity of $3.3^{+0.4}_{-0.3}\%$, post-selected on atom survival, arising primarily from failure to detect a cavity photon during the detection window. For Rydberg-state detection, background dark counts contribute an infidelity of $1.6^{+0.2}_{-0.1}\%$. Including the estimated error due to $T_1$ decay during detection yields a total Rydberg-state detection infidelity of $4.6^{+0.2}_{-0.1}\%$; see Supplemental Material~\cite{supplemental}.

We demonstrate nondestructive readout by measuring Rabi oscillations between the ground and Rydberg states while preserving atom survival; see Fig.\ \ref{fig:CavDetect}(d). We measure an average atom loss of $6.5(7)\%$, with the most loss occurring when the Rydberg population is high and the tweezer remains off throughout detection. Losses are consistent with release-and-recapture loss at our measured atom temperature of $12(2)~\mu\mathrm{K}$ and can be further reduced through improved cooling or shorter tweezer-off times. We note that this experiment was performed with a single atom coupled to the cavity, but these results could be extended to arrays by using local addressing beams~\cite{mcmvladan,Lee2026Rapid}.

A second unique capability of the combined platform is the simultaneous control of finite-range Rydberg interactions and collective coupling of many emitters to a shared optical mode. We independently tune these interactions through the interatomic separation and atomic positions within the cavity standing wave, respectively, realizing blockaded and unblockaded Rydberg dynamics together with superradiant and subradiant optical responses.

The collective optical response is controlled by the atoms’ positions within the cavity standing wave, with the cavity coupling for atom $i$ at position $x_i$ being $g_i=g_0\cos(kx_i)$, where $k=2\pi/\lambda$ is the wavenumber of the cavity mode at a wavelength $\lambda=780$ nm. For two atoms, the cavity couples to the collective dipole $g_1\sigma_1^-+g_2\sigma_2^-$, where $\sigma_i^-$ is the atomic lowering operator for the $D_2$ transition.   Positioning the atoms at same-sign $(g_1=g_2)$ or opposite-sign $(g_1=-g_2)$ antinodes produces constructive (superradiant) or destructive (subradiant) interference, respectively~\cite{Dicke54, yan2023superradiant}.

The Rydberg interaction $V(R)$ is controlled, instead, by the interatomic separation $R$. For atom-atom interaction strength $V(R)\gg\Omega$, Rydberg blockade suppresses coupling to $\ket{rr}$, producing oscillations between the $\ket{gg}$ and $\ket{W}=(\ket{gr}+\ket{rg})/\sqrt{2}$ two-atom states when optically driven.  The Rabi frequency for this two-atom blockaded oscillation is $\sqrt{2}$ times larger than the single-atom Rabi frequency.  For $V(R)\ll\Omega$, the two-atom $\ket{rr}$ state becomes accessible, and the Rabi frequency returns to the single-atom value~\cite{Saffman2010_rev}.

\begin{figure}[t]
    \centering
    \includegraphics[width=1\linewidth]{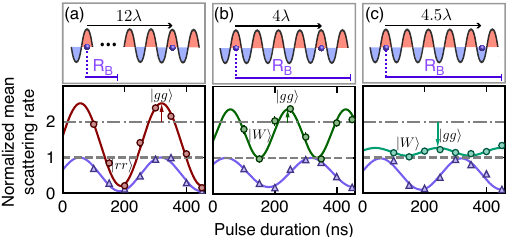}
    \caption{Superradiant and subradiant cavity emission with Rydberg blockade. Single-atom Rydberg Rabi oscillations (purple triangles) are compared with two-atom Rabi oscillations for three configurations. The mean scattering rate is normalized to the single-ground-state-atom rate of $0.42(2)$~$\mu \text{s}^{-1}$; gray dash-dotted and dashed lines denote $1 \times$ and $2 \times$ the single-atom emission levels, respectively. (a) Unblockaded, superradiant configuration with $R=12\, \lambda=9.4~\mu\mathrm{m}>R_B$, where $R_B$ is the blockade radius, and $g_1=g_2$. (b) Blockaded, superradiant configuration with $R=4\, \lambda=3.1~\mu\mathrm{m}<R_B$ and $g_1=g_2$. (c) Blockaded, subradiant configuration with $R=4.5\, \lambda=3.5~\mu\mathrm{m}<R_B$ and $g_1=-g_2$. Solid lines are sinusoidal fits to the mean scattering rate, reflecting oscillations in the probabilities of discrete scattering outcomes. Error bars denote $68\%$ confidence intervals. 
    }
    \label{fig:super_sub_block}
\end{figure}

We probe these two forms of collective behavior by monitoring cavity emission following coherent Rydberg excitation of pairs of atoms in the cavity. Because ground-state atoms scatter collectively into the cavity while Rydberg atoms remain dark, the photon counts simultaneously probe the Rydberg dynamics and the collective optical response. We show three two-atom configurations  with different Rydberg interaction strengths and relative sign of the atom–cavity coupling (Fig.~\ref{fig:super_sub_block}). For these measurements, we modify the detection sequence of Fig.\ \ref{fig:CavDetect} to improve atom localization within the cavity standing wave: the tweezers are not adiabatically ramped down from full depth, are switched off only during Rydberg excitation, and are restored before a fixed $5~\mu\mathrm{s}$ cavity-detection window.

At a large atom-atom separation of $12\, \lambda$, the weak Rydberg interaction, $V(R)\approx2\pi\times46~\mathrm{kHz}\ll\Omega=2\pi\times3.8~\mathrm{MHz}$, permits excitation to $\ket{rr}$, while same-sign cavity coupling $(g_1=g_2)$ produces a superradiant response; see Fig.\ \ref{fig:super_sub_block}(a). At the $\ket{gg}$ maxima, the mean scattering rate exceeds the incoherent two-atom expectation by a factor of $\mathcal S = 1.3(1)$, whereas excitation to the doubly excited Rydberg states $\ket{rr}$ shows no emission into the cavity.

Reducing the atom-atom separation to $4\,\lambda$ increases the interaction to $V(R)\approx2\pi\times34~\mathrm{MHz}\gg\Omega$, producing blockaded $\ket{gg}\leftrightarrow\ket{W}$ oscillations. The $\ket{gg}$ state retains the same-sign, superradiant cavity response with enhancement factor $\mathcal S=1.2(1)$, while the singly excited $\ket{W}$ state scatters as a single ground-state atom. Finally, shifting to $4.5\,\lambda$ preserves the blockade ($V(R) \approx 2\pi\times17~\mathrm{MHz}\gg\Omega$) but reverses the relative coupling $(g_1=-g_2)$, switching the $\ket{gg}$ response from superradiant to subradiant and reducing the mean scattering rate to $\mathcal S=0.6(1)$ times the incoherent two-atom sum.
The observed degree of superradiant enhancement (subradiant suppression) of collective light scattering is below (above) the ideal value $\mathcal S = 2$ ($\mathcal S = 0$), owing to imperfect atom localization within the cavity and probe standing waves and to jumps between magnetic sublevels during the probe period; see Supplemental Material for details~\cite{supplemental}.  

We have demonstrated a cavity–Rydberg platform combining coherent Rydberg control with single-atom strong coupling to an optical cavity. Enabled by the strong suppression of electric field fluctuations in our cavity, we demonstrate fast, nondestructive ground–Rydberg detection within the Rydberg-state lifetime. By independently controlling interatomic separation and cavity coupling, we observed blockaded and unblockaded Rydberg dynamics with superradiant and subradiant optical responses.
Looking forward, reduced Rydberg-excitation laser phase noise would increase the coherence time of the ground-Rydberg transition, and ground-state motional cooling would further improve atom localization and coherent coupling to the cavity.  We note that while the present experiments were performed largely on single atoms and atom pairs, the methods demonstrated here can be readily applied to the larger atom arrays that have been previously introduced into strongly coupled optical cavities~\cite{ye26controlling,Wang2025CavityQED,DeSantis2026CavityRydberg,ho2025optomechanical,yu2026stochastic}.

Our work provides a foundation for networks of Rydberg microprocessors connected by light, where photonic links distribute entanglement generated via Rydberg gates between processors and enable quantum operations to be teleported across network nodes~\cite{young2022architecture,covey_quantum_2023, sinclair2025link,li2024interconnect}. Additionally, the techniques demonstrated here enable measurement and feed-forward during Rydberg many-body dynamics, measurement-assisted quantum optimization~\cite{ferguson2021_MBVQE}, and the exploration of phenomena such as measurement-induced phase transitions in interacting Ising systems~\cite{chaki2026_notrans}. Finally, the platform opens new directions for quantum simulation, including the exploration of exotic phases of matter~\cite{Roman-Roche2025_DI}, competing interaction length scales~\cite{hosseinabadi2026_RydQED}, and geometric frustration~\cite{liang2026_DI}.

We acknowledge helpful discussions with Johannes Zeiher, Jacopo De Santis, Bal\'azs Dura-Kov\'acs, Mehmet \"Onc\"u, Adrien Bouscal, Sepehr Ebadi, and Jacob Covey. We acknowledge support from the AFOSR (Grant No.\ FA9550-1910328), from ARO through the MURI program (Grant No.\ W911NF-20-1-0136) and the DURIP program (Grant No.\ W911NF2310244), from DARPA (Grant No.\ W911NF2010090), from the NSF through the CIQC QLCI Program (Grant No.\ OMA-2016245) and the MRI program (Grant No.\ OMA-2216201), and from the U.S. Department of Energy, Office of Science, National Quantum Information Science Research Centers, Quantum Systems Accelerator (Award No. DE-SCL0000121).
J.H.\ acknowledges support from the Department of Defense through the National Defense Science and Engineering Graduate (NDSEG) Fellowship Program. N.B.V.\ acknowledges support from the Miller Institute for Basic Research in Science.

\bibliography{main}
\clearpage

\onecolumngrid
\pagebreak


\section*{Supplementary material}

# Supplemental Material for
# A Rydberg atom array coupled to a high-cooperativity optical cavity

Tai Xiang,[1,2] Yue-Hui Lu,[1,2] Tsai-Chen Lee,[1,2] Jacquelyn Ho,[1,2] Michael Lan,[1,2] Zhenjie Yan,[1,3] Harry Levine,[1,2,4] Nathaniel B. Vilas,[1,2] and Dan M. Stamper-Kurn[1,2,5]

[1]*Department of Physics, University of California, Berkeley, California 94720*
[2]*Challenge Institute for Quantum Computation, University of California, Berkeley, California 94720*
[3]*Department of Physics, Columbia University, New York, NY 10027*
[4]*Oratomic, Pasadena, California, 91125*
[5]*Materials Sciences Division, Lawrence Berkeley National Laboratory, Berkeley, California 94720*

## I. RYDBERG EXCITATION

Here we describe the state preparation and Rydberg excitation protocol in greater detail. We apply a magnetic field of $B \simeq 3$ G along the Rydberg excitation axis (see main text, Fig. 1), defining the quantization axis, and optically pump the atoms into the stretched ground state $|5S_{1/2}, F = 2, m_F = 2\rangle$. Prior to Rydberg excitation, we adiabatically ramp the tweezer trap depth from 510 $\mu$K to 51 $\mu$K over $\sim$160 $\mu$s, thereby reducing the atom temperature from 27(3) $\mu$K to 13(2) $\mu$K. We then turn off the tweezers for up to 10 $\mu$s during Rydberg operations. Note that the adiabatic tweezer rampdown is not used for the results presented in Fig. 5 of the main text.

Rydberg excitation is driven through the two-photon transition $|5S_{1/2}, F = 2, m_F = 2\rangle \to |6P_{3/2}, F = 3, m_F = 3\rangle \to |nS_{1/2}, m_J = 1/2, m_I = 3/2\rangle$, using $\sigma^+$-polarized lower-leg light 420 nm in wavelength and $\sigma^-$-polarized upper-leg light 1014–1018 nm in wavelength, depending on the principal quantum number $n$. The lower-leg beam is focused to a waist of 59(1) $\mu$m. The upper-leg beam has elliptical waists of 49(1) $\mu$m along the $z$ (transverse) axis and 92(3) $\mu$m along the $x$ (cavity) axis. The lower-leg laser is frequency stabilized to the $|5S_{1/2}, F = 2\rangle \to |6P_{3/2}, F = 3\rangle$ transition using an atomic vapor cell, while the upper-leg laser is stabilized to an ultralow-expansion reference cavity. For $n$=53, we achieve two-photon Rabi frequencies up to $\Omega = 2\pi \times 5$ MHz.

When exciting Rydberg atoms, we begin applying the 1014-1018-nm-wavelength $|6P_{3/2}\rangle \to |nS_{1/2}\rangle$ light $\gtrsim$100 $\mu$s prior to Rydberg excitation, and then pulse on the 420-nm-wavelength $|5S_{1/2}\rangle \to |6P_{3/2}\rangle$ light. We use an acousto-optic modulator with a rise time of $\sim$ 60 ns to control the duration, detuning, and phase of the lower-leg light driving the $|5S_{1/2}\rangle \to |6P_{3/2}\rangle$ transition. For the Ramsey coherence measurements in Fig. 3 of the main text, we detune the excitation light from the two-photon Rydberg resonance, producing a relative phase between the two $\pi/2$ pulses, referenced to the atomic transition, that advances linearly with the hold time.

## II. ZONED ATOM-ARRAY ARCHITECTURE WITHIN A STRONGLY COUPLED OPTICAL CAVITY

For several of the measurements described in the main text, we use a zoned tweezer array architecture. Specifically, we load atom arrays initially with large atom number, with as many as $N = 15$ atoms in a one-dimensional array, and separate atoms spatially between two storage zones and a Rydberg interaction zone. We use this architecture to perform spectroscopy in which the final state measurement is performed by imaging atomic fluorescence. For such spectroscopy, we move one or more atoms from the storage zone into the interaction zone. Following state preparation, we selectively illuminate atoms only in the interaction zone with pulses of light that drive the two-photon Rydberg excitation. We then measure the presence of trapped ground-state atoms by fluorescence imaging, with atoms in Rydberg states being detected indirectly as lost atoms. After detection, we move atoms from the storage zones to the interaction zone to perform the next measurement. In this manner, within a single tweezer-array loading sequence, we are able to perform multiple measurements, increasing our data rate as we perform Stark-shift spectroscopy, observe Rabi oscillations, and measure Ramsey coherence times. This approach exemplifies how a zoned quantum computing architecture may be realized with large tweezer arrays placed either entirely or in part within a strongly coupled optical cavity.

For experiments involving cavity fluorescence, such as the direct low-loss detection of atoms in the ground-Rydberg basis and the coherent oscillation of atom pairs between collectively scattering and non-collectively scattering states, we place only a single atom or a single atom pair in the cavity. We avoid maintaining additional atoms in an intra-cavity storage zone so that those atoms do not interfere with cavity fluorescence. In future studies, one may adopt the optical control of atomic excited states, as demonstrated in Ref. [S1], to enable mid-circuit measurements of single-

atom or many-atom states through cavity interactions without having such measurements disturb or be disturbed by unmeasured atoms within an intra-cavity array.

## III. UV PROCEDURE

We use ultraviolet (UV) illumination to remove charges on surfaces near the atoms. Two UV LEDs 375 nm in wavelength (Thorlabs M375L4 and Shark OTLH-0280-UV), positioned on opposite sides of the vacuum chamber that houses the cavity, illuminate the cavity mirrors and surrounding mount surfaces. We verify illumination of the cavity mirrors by imaging the cavity mirrors with the UV light onto a camera.

We perform UV cleaning with all piezoelectric cavity actuators grounded. During the cleaning procedure, we continuously illuminate the cavity assembly with UV light while monitoring the Rydberg resonance. The UV illumination is stopped once the resonance stabilizes to a constant frequency (see Fig. S1), indicating that the local electric-field environment has reached a steady state. From Rydberg state spectroscopy, we find that this steady state typically corresponds to an electric field magnitude $|E| < 0.4$ V/cm at the atoms. The electric field can be further zeroed along the $x$ and $z$ axes by applying bias voltages to the piezo actuators.

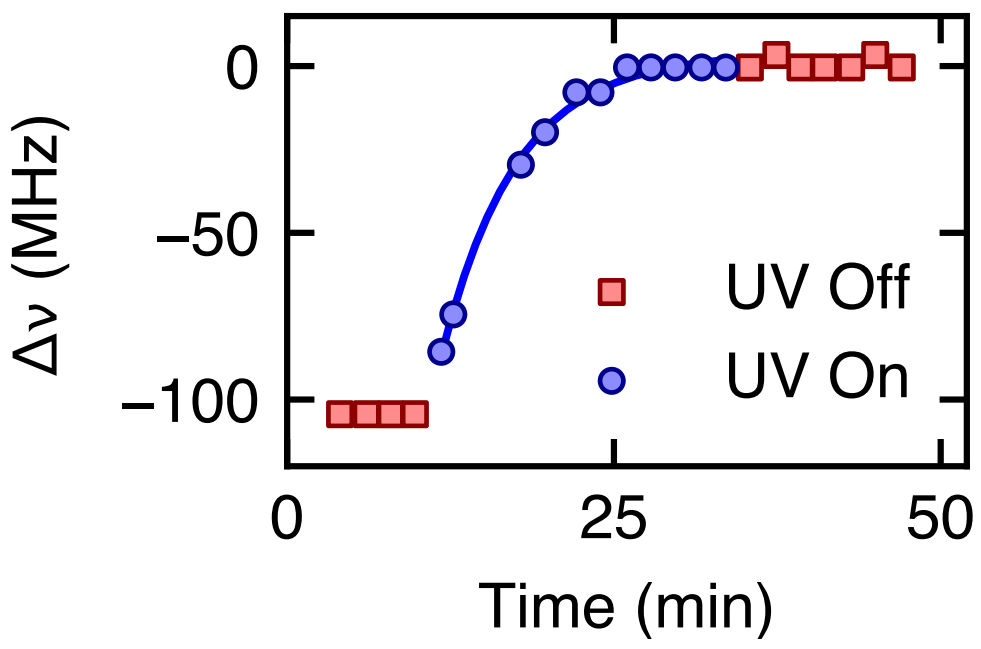


Figure S1. Rydberg resonance shift $\Delta\nu$ under UV illumination (blue circles) and when UV is switched off (red squares). Solid blue line denotes an exponential fit to representative data, yielding a $1/e$ decay time of 5.7(4) minutes. This timescale varies with experimental conditions.

If the UV illumination is applied during active cavity stabilization (i.e. when voltages are being applied to the actuators), we observe substantial drifts of the Rydberg resonance over time. We hypothesize that this slow response in the presence of UV illumination is due to time-dependent redistribution of surface charge in the presence of the actuator-generated electric fields. Although the compensation scheme described in the main text suppresses the actuator-induced electric field at the atomic position, changes in the actuator voltages modify the spatially varying electric field in the vacuum chamber. UV illumination in the presence of these changing fields can redistribute surface charge, potentially producing time-dependent electric fields at the atoms. We therefore perform UV cleaning only with the cavity actuators grounded and turn off the UV light source before engaging active cavity stabilization and performing experimental measurements.

We typically find that the Rydberg resonance frequency is stable for timescales of a few days or more during normal operation, after which time UV cleaning needs to be reapplied.

## IV. CALIBRATING AND COMPENSATING ELECTRIC FIELDS FROM THE PIEZOELECTRIC ACTUATORS

We discuss the electric field compensation technique in greater detail. As illustrated in Fig. 1(c) of the main text, the cavity mirror positions are controlled by two shear-stack piezoelectric actuators (Noliac NAC2402-H2.3). Each actuator consists of an inner plate electrode and two outer plate electrodes shorted together. Using the manufacturer-

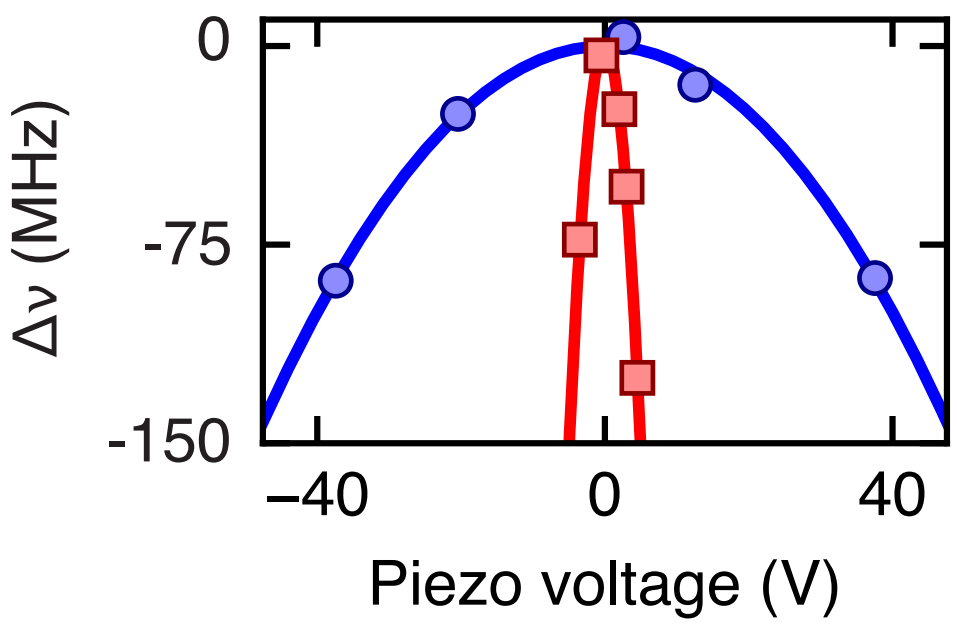


Figure S2. Rydberg resonance shift as a function of voltage applied simultaneously to both inner (blue circles) or both outer (red squares) plate electrodes.

specified coefficient $\alpha_p$, the displacement of cavity mirror $i$ ($i = 1, 2$) is given by

$$\Delta x_i = \alpha_p (V_{\mathrm{inner},i} - V_{\mathrm{outer},i}), \qquad \alpha_p = 9.4\ \mathrm{nm/V}, \tag{S1}$$

where $V_{\mathrm{inner},i}$ and $V_{\mathrm{outer},i}$ are the voltages applied to the inner and outer electrodes of actuator $i$, respectively. The corresponding cavity-length change is $\Delta L = \Delta x_1 + \Delta x_2$. Throughout Fig. 2 of the main text, the cavity-length displacement $\Delta L$ plotted on the horizontal axis is calculated from the applied electrode voltages using $\alpha_p$, rather than through direct measurements of the cavity length. The electric field at the location of the atoms depends linearly on the four electrode voltages. By symmetry of the electrode geometry, the applied electric field is predominantly along the $x$ and $z$ axes.

We lock the cavity length to a Pound-Drever-Hall error signal using a single voltage $V_{\mathrm{ctrl}}$ applied to both inner plate electrodes, so that $V_{\mathrm{inner},1} = V_{\mathrm{inner},2} = V_{\mathrm{ctrl}}$. Additionally, scaled copies of $V_{\mathrm{ctrl}}$ are fed forward to the two outer electrodes,

$$V_{\mathrm{outer},i} = c_i V_{\mathrm{ctrl}} + V_{\mathrm{bias},i}, \tag{S2}$$

where the feed-forward coefficients $c_1$ and $c_2$ are chosen to cancel the electric field at the location of the atoms as $V_{\mathrm{ctrl}}$ varies to stabilize the cavity length. The bias voltages $V_{\mathrm{bias},i}$ are used to apply an additional static bias electric field to the atoms, e.g. to cancel small residual fields remaining after UV cleaning. With these definitions, $\Delta L = (9.4\ \mathrm{nm/V}) \times (2 - c_1 - c_2) V_{\mathrm{ctrl}} \equiv c_L V_{\mathrm{ctrl}}$, where $c_L$ is the resulting cavity-length tuning coefficient. Note that a constant length offset due to the bias voltages has been absorbed into the definition of $\Delta L$.

To obtain an initial estimate for the coefficients $c_1$ and $c_2$, we independently measure the electric-field sensitivity of the inner and outer plate electrodes along each axis. For each electrode, we vary the applied voltage, measure the shift in the Rydberg resonance, and fit the resultant frequency shifts to a parabola of the form $\delta\nu(V) = A(V - V_0)^2 + \delta\nu_0$ where $V$ is the applied voltage, $V_0$ and $\delta\nu_0$ are fit offsets, and the curvature $A$ is determined by the electric-field change per unit applied voltage; see Fig. S2. Comparing the fitted curvatures for the inner and outer plate electrodes provides an initial estimate of $c_1$ and $c_2$. We then empirically refine the compensation by minimizing the Rydberg resonance shift observed at large values of $V_{\mathrm{ctrl}}$ in each axis. Because adjustments that minimize the residual sensitivity along one axis can modify the sensitivity along the orthogonal axis, we alternate between $x$- and $z$-axis measurements and iteratively tune the two coefficients until the residual sensitivity along both controlled axes is sufficiently small.

Upon applying the compensation, we measure residual sensitivities of 0.085(8) and 0.066(17) mV/cm/nm along the $z$ and $x$ axis, respectively, limited by the calibration accuracy of the plate voltage ratios and can, in principle, be further reduced.

## V. RESPONSE TO ELECTRIC FIELD NOISE

Here, we consider the effect of shot-to-shot Gaussian electric-field fluctuations on the $T_2^*$ coherence time of ground–Rydberg superpositions. The dc Stark shift of the ground–Rydberg transition frequency is $\Delta\nu = -\frac{1}{2}\alpha E_b^2$, where $\alpha$ is the polarizability expressed in frequency units and $E_b$ is the applied bias field. Electric-field fluctuations therefore produce fluctuations in the transition frequency. We model the total electric field as $E = E_b + \delta E$, where $\delta E$ represents quasi-static, shot-to-shot electric field fluctuations drawn from a zero-mean Gaussian random variable with standard

deviation $\sigma_E$. The ensemble-averaged Ramsey signal at time $t$ is

$$P(t) = \mathrm{Re}\left[\langle e^{-i\omega t}\rangle\right] \tag{S3}$$

$$= \mathrm{Re}\left[\frac{1}{\sqrt{2\pi}\sigma_E}\int_{-\infty}^{\infty} e^{i\pi\alpha(E_b+\delta E)^2 t} e^{-\delta E^2/(2\sigma_E^2)}\, d(\delta E)\right] \tag{S4}$$

$$= \mathrm{Re}\left[e^{-i\omega_0 t}\frac{1}{\sqrt{1-2i\pi\alpha\sigma_E^2 t}}\exp\left(-\frac{2\pi^2\alpha^2 E_b^2\sigma_E^2 t^2}{1-2i\pi\alpha\sigma_E^2 t}\right)\right], \tag{S5}$$

where $\omega_0 = 2\pi\Delta\nu$ is the angular-frequency shift induced by the dc Stark effect, which can be removed by recalibrating the Rydberg resonance frequency.

We consider two limits. First, in the limit $E_b \gg \sigma_E$, the Stark shift can be linearized about $E_b$, yielding Gaussian-distributed angular frequency fluctuations $\delta\omega = -2\pi\alpha E_b \delta E$ with standard deviation $\sigma_\omega = 2\pi\alpha E_b\sigma_E$. The Ramsey envelope function then becomes

$$\mathrm{Re}\left[\frac{1}{\sqrt{2\pi}\sigma_\omega}\int_{-\infty}^{\infty} e^{-i\delta\omega t} e^{-\delta\omega^2/(2\sigma_\omega^2)}\, d(\delta\omega)\right] = e^{-\sigma_\omega^2 t^2/2}, \tag{S6}$$

corresponding to a Gaussian coherence envelope with

$$T_{2,E}^* = \frac{1}{\sqrt{2}\pi\alpha E_b\sigma_E}. \tag{S7}$$

Second, at zero bias $E_b \to 0$, the first-order sensitivity to electric-field fluctuations vanishes, and the coherence envelope instead exhibits a power-law decay, asymptotically scaling as $t^{-1/2}$.

Following electric-field compensation, we estimate residual bias fields of $|E_b| \lesssim 10$ mV/cm along each controlled axis. Electric-field fluctuations along these axes therefore contribute negligibly to the zero-bias Ramsey decay. We attribute the observed 1.1 $\mu$s coherence time primarily to laser phase noise, although residual electric-field noise along the uncontrolled axis may also contribute. We thus model the Ramsey fringe envelope as a product of independent Gaussian decay contributions from background dephasing and electric-field fluctuations. In the limit $E_b \gg \sigma_E$, the resulting coherence time is

$$T_2^*(E_b) = \left[\frac{1}{(T_{2,\mathrm{bg}}^*)^2} + 2\pi^2\alpha^2 E_b^2\sigma_E^2\right]^{-1/2}, \tag{S8}$$

where $T_{2,\mathrm{bg}}^*$ describes bias-independent dephasing. We fit the measured bias-dependent coherence times to this expression to extract the rms electric-field noise amplitude $\sigma_E$.

## VI. RAMSEY COHERENCE TIME FITTING WITH LINEAR DRIFT

We observe a slow drift of the electric field at the atoms when the piezoelectric electrodes are held at a constant voltage or grounded. This slow drift may be caused by dipolar relaxation of the piezoelectric material [S2], a phenomenon known as piezoelectric creep. This effect is most pronounced when the Rydberg resonance is deliberately biased away from the zero-field point, where the Rydberg-state energy becomes first-order sensitive to electric-field fluctuations; see Fig. S3.

To account for this systematic drift, we incorporate a time-dependent detuning into the Ramsey model, $\Delta(s) = 2\pi \times (f_0 + ms)$, where $s$ denotes the experimental time when a shot is acquired, $f_0$ is the initial detuning, and $m$ is the detuning drift rate. For a Ramsey free-evolution time $\tau$, the accumulated phase is therefore $\Delta(s)\tau + \phi_0$. We model the probability of detecting the atom in the measured state as

$$P(\tau, s; \boldsymbol{\theta}) = C - A\exp\left[-\left(\frac{\tau}{T_2^*}\right)^2\right]\cos\left[2\pi(f_0+ms)\tau + \phi_0\right], \qquad \boldsymbol{\theta} = (C, A, T_2^*, \phi_0, f_0, m). \tag{S9}$$

Here, $C$ is the offset, $A$ is the Ramsey-fringe amplitude, $T_2^*$ is the Ramsey coherence time, and $\phi_0$ is an overall phase offset. Rather than averaging repeated measurements at each Ramsey evolution time, we retain the binary outcome

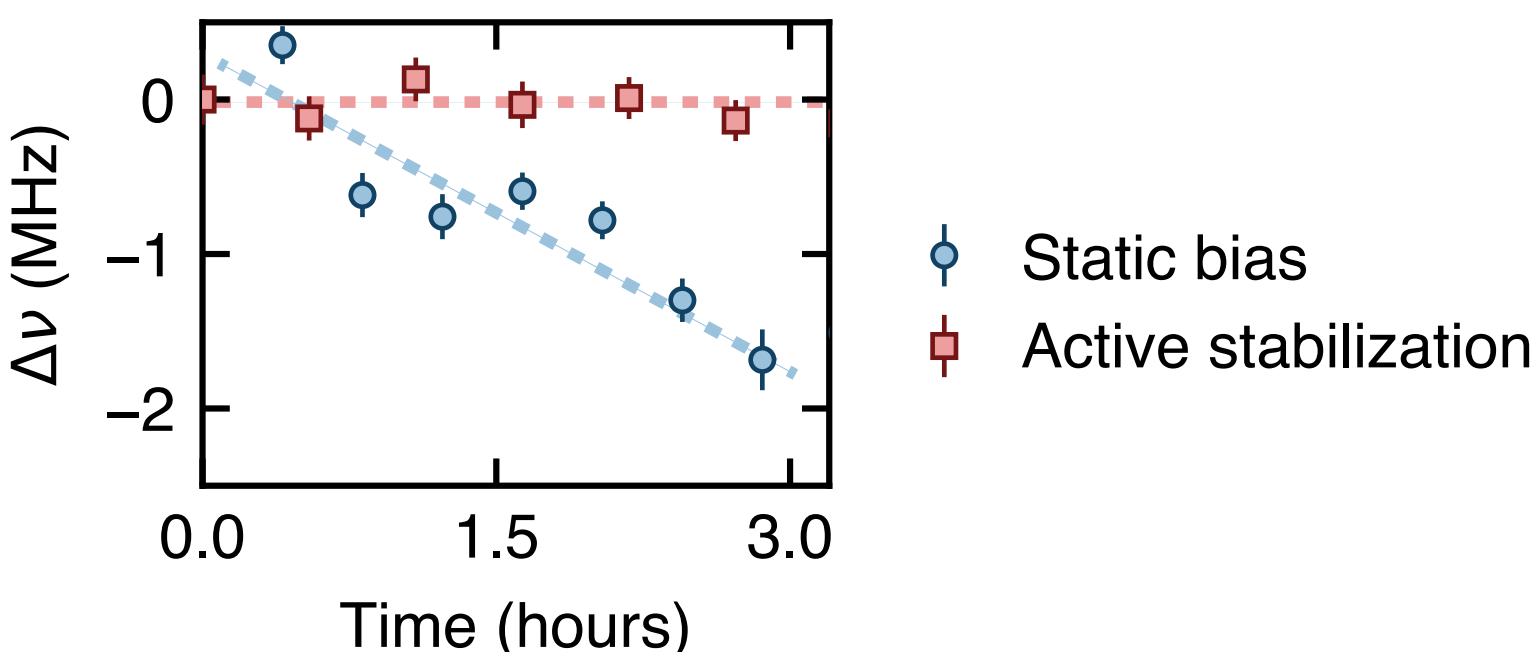


Figure S3. We continuously measure the Rydberg resonance and group experimental shots into time bins of 200 shots. The fitted Lorentzian resonance center exhibits a linear drift without cavity-length stabilization (red squares), while no appreciable drift is observed with active cavity stabilization engaged (blue circles). Both datasets are acquired at an electric-field bias of 1.0 V/cm. Error bars denote $1\sigma$ fit uncertainties, and dashed lines are guides to the eye.

and acquisition time of each individual shot. Assuming independent Bernoulli-distributed measurement outcomes, the likelihood of the full dataset given experimental shot $i$ with binary outcome $y_i \in \{0, 1\}$ is

$$\mathcal{L}(\boldsymbol{\theta}) = \prod_i P_i^{y_i}(1 - P_i)^{1-y_i}, \tag{S10}$$

and we determine the fit parameters by minimizing the negative log-likelihood for the entire experimental dataset,

$$-\log \mathcal{L}(\boldsymbol{\theta}) = -\sum_i \left[y_i \log P_i + (1 - y_i) \log(1 - P_i)\right]. \tag{S11}$$

This shot-resolved analysis accounts for the deterministic phase evolution caused by the slow resonance drift, allowing $T_2^*$ to be extracted independently of the systematic frequency drift.

We do not observe a similar drift of the Rydberg resonance when the cavity length is actively stabilized using feedback to a Pound-Drever-Hall error signal, as shown in Fig. S3. We hypothesize that cavity-length feedback suppresses the electric-field changes associated with slow relaxation of the piezoelectric material. When the actuators are grounded following a change in applied voltage, the piezoelectric polarization and strain slowly relax toward equilibrium, producing both mechanical creep and a time-dependent electric field at the atoms. Under active stabilization, the feedback continuously adjusts the actuator voltage to maintain a fixed cavity length, thereby fixing the polarization and the strain in the piezo material as well.

## VII. NONDESTRUCTIVE RYDBERG STATE DETECTION SEQUENCE

Here, we describe the single-atom, nondestructive ground–Rydberg state detection protocol (Fig. 4 of the main text) in greater detail. Immediately following Rydberg excitation, described above in Section I of the Supplemental Material, the atom is probed with a retro-reflected light field, 780 nm in wavelength and linearly polarized along $z$, that induces Purcell-enhanced fluorescence. The probe light is resonant with the optical cavity and detuned by $\Delta_{\rm pa} = -2\pi \times 55$ MHz from atomic resonance, and it has a Rabi frequency of $\Omega_{\rm p} = 2\pi \times 27$ MHz (referenced to the $|F = 2, m_F = \pm 2\rangle \to |F' = 3, m'_F = \pm 3\rangle$ cycling transition). Following a fixed probing window of 7.5 $\mu$s, we illuminate the atom with a resonant depumping light field 1015 nm in wavelength with Rabi frequency $\Omega_{\rm dp} = 2\pi \times 495$ MHz. Atoms that are successfully transferred to the Rydberg state are depumped to the intermediate state before decaying back down to the ground state, whereas ground state atoms are not affected by the depumper light. We find that 2.5 $\mu$s of depumping time is sufficient to saturate the survival of atoms excited to the Rydberg state; see Fig. S4(b). We summarize the pulses applied in the protocol in Fig. S4.

## VIII. RYDBERG LIFETIME CONTRIBUTION TO MEASUREMENT INFIDELITY

The measured false-positive rate shown in the histogram in Fig. 4(c) does not include false positives arising from decay of the Rydberg state during detection. We account for this contribution using theoretically calculated transition

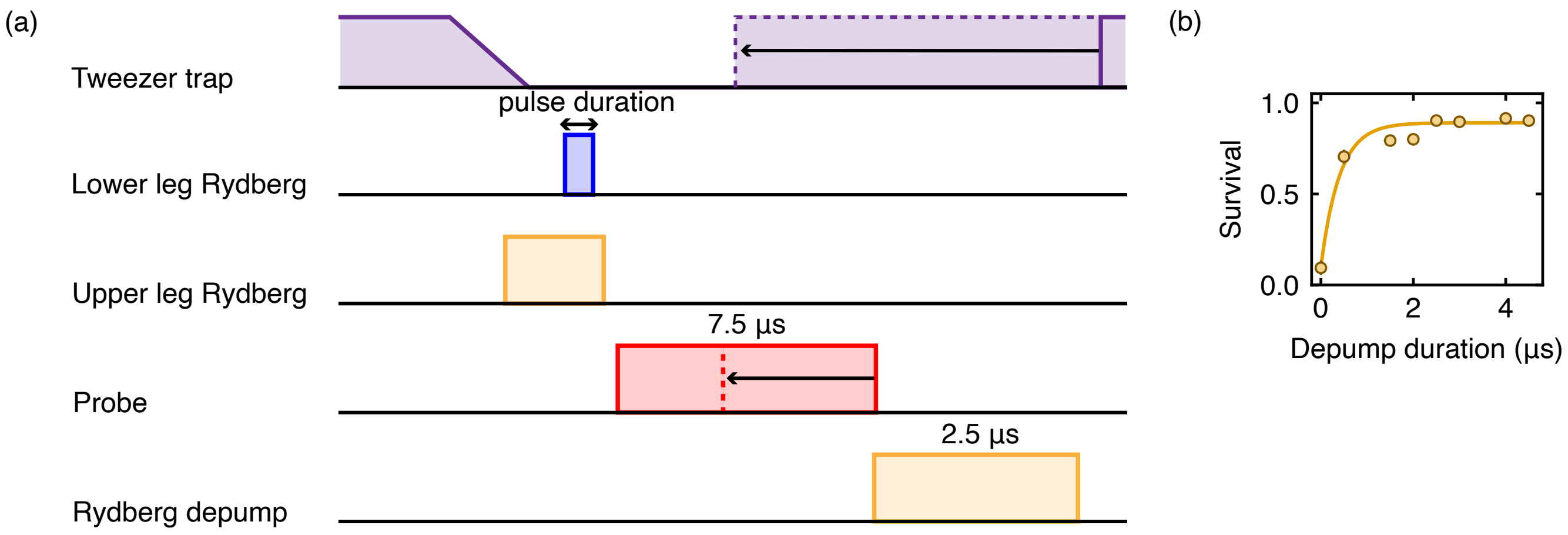


Figure S4. (a) Summary of the pulse sequence applied for nondestructive ground-Rydberg state detection. Dashed lines correspond to portions of the sequence that may be adaptively controlled in response to a detected photon. (b) Timescale for depumping atoms from the Rydberg state back into the ground state.

rates from $|53S_{1/2}\rangle$ [S3], including both spontaneous emission and blackbody-radiation-induced transitions. Because only atoms in the $|5S_{1/2}, F = 2\rangle$ manifold couple to the cavity probe, the relevant timescale is not simply the lifetime of the $|53S_{1/2}\rangle$ state, but also includes the time required for the atom to cascade through intermediate states before reaching the ground-state manifold. A repump field that returns population in $|5S_{1/2},\ F = 1\rangle$ to $|5S_{1/2},\ F = 2\rangle$ is on during detection; for simplicity, we assume repumping to be instantaneous and therefore treat all population reaching $|5S_{1/2}\rangle$ as bright. We calculate the effective decay time by accounting for all the allowed transitions out of $|53S_{1/2}\rangle$ and the subsequent cascade through intermediate states. We obtain a mean decay time $T_{\rm decay} = 173\ \mu$s for an atom initially prepared in $|53S_{1/2}\rangle$ to reach the ground-state manifold. We approximate the distribution of arrival times in the ground-state manifold as an exponential, and we therefore describe the effective decay into the ground state by

$$f_{\rm decay}(t) = \Gamma_{\rm decay} e^{-\Gamma_{\rm decay} t}, \qquad \Gamma_{\rm decay} = T_{\rm decay}^{-1}. \tag{S12}$$

For an atom that reaches the ground state at time $t$, the remaining time for photon detection is $T_{\rm probe} - t$. Assuming that detected cavity photons follow a Poisson process, the probability of detecting at least one photon during this interval is $P_{\rm det}(t) = 1 - e^{-R_{\rm det}(T_{\rm probe}-t)}$, where $R_{\rm det}$ is the detected cavity-photon rate, which we measure to be $0.43(2)\ \mu s^{-1}$. The resulting decay-induced false-positive probability is

$$\epsilon_{\rm decay} = \int_0^{T_{\rm probe}} \Gamma_{\rm decay} e^{-\Gamma_{\rm decay} t} \left[1 - e^{-R_{\rm det}(T_{\rm probe}-t)}\right] dt, \tag{S13}$$

which, for our experimental parameters, gives an additional false-positive contribution of 3.0%. This brings the total false-positive rate to 4.6%.

## IX. REDUCTION IN SUPERRADIANCE

We further discuss the reduction in superradiant fluorescence enhancement, relative to the ideal theoretical value, observed in Fig. 5 of the main text. Superradiant emission relies on the coherent addition of scattered fields from indistinguishable atomic dipoles. In our experiment, finite atomic temperature and the ballistic expansion that occurs while the tweezers are switched off during Rydberg excitation reduce the localization of the atoms within the cavity and retro-reflected probe standing waves, producing shot-to-shot fluctuations in the magnitude and phase of the emitted field. Additionally, mixing among Zeeman sublevels, which introduces state-dependent transition strengths and detunings, further reduces the enhancement.

### A. Finite atom localization

We first consider the effect of imperfect atomic localization and motional dynamics on the observed superradiant enhancement. During the experimental sequence, the optical tweezers are switched off for a duration $t_r = 670$ ns before being restored for cavity fluorescence detection. We model the atomic motion during release and recapture using the evolution of the position and momentum variances, $V_x = \langle x^2 \rangle - \langle x \rangle^2$ and $V_p = \langle p^2 \rangle - \langle p \rangle^2$, and their covariance $C_{xp} = \frac{1}{2}\langle \{x, p\} \rangle - \langle x \rangle \langle p \rangle$, where $x$ and $p$ are the position and momentum operators. For a harmonic oscillator with time-dependent trap frequency $\omega(t)$, starting from any initial Gaussian state, the moments obey the closed set of equations

$$\begin{aligned} \partial_t V_x &= \frac{2}{m} C_{xp}, \\ \partial_t V_p &= -2m\omega^2(t) C_{xp}, \\ \partial_t C_{xp} &= \frac{V_p}{m} - m\omega^2(t) V_x. \end{aligned} \tag{S14}$$

We initialize the atoms in thermal equilibrium at temperature $T$ in a harmonic trap with radial frequency $\omega_r$, with $C_{xp}(0) = 0$. In the high-temperature limit, $V_x(0) = k_B T/(m\omega_r^2)$, $V_p(0) = mk_B T$. During the trap-off interval, $\omega(t) = 0$, and the position variance increases due to free expansion. Upon recapture, $\omega(t) = \omega_r$, and the position variance undergoes oscillations at twice the radial trap frequency during the 5 $\mu$s probing period.

To determine the resulting reduction in collective emission due to the aforementioned dynamics, following Ref. [S4], we write the position-dependent scattering amplitude of atom $i$ as $\eta_i = \eta_0 q_i$, where

$$q_i = \cos(kx_i)\cos(ky_i), \tag{S15}$$

$k = 2\pi/\lambda$, and $\eta_0$ is the single-atom scattering amplitude at the antinodes of both the cavity and transverse-probe standing waves. For two identically distributed, independently moving atoms positioned at equivalent antinodes, the superradiant enhancement relative to the incoherent sum of single-atom scattering rates is

$$\mathcal{S}_{\text{pos}}(t) = 1 + \mathcal{D}(t), \qquad \mathcal{D}(t) = \frac{\langle q(t) \rangle^2}{\langle q^2(t) \rangle}, \tag{S16}$$

where $\mathcal{D}(t)$ is the motional Debye–Waller factor. Perfect localization gives $\mathcal{D} = 1$ and $\mathcal{S}_{\text{pos}} = 2$.

For Gaussian position distributions centered at the standing-wave antinodes, with independent and equal variances $V_x(t) = V_y(t)$ along the cavity and probe axes,

$$\langle q(t) \rangle = e^{-k^2 V_x(t)}, \qquad \langle q^2(t) \rangle = \left[ \frac{1 + e^{-2k^2 V_x(t)}}{2} \right]^2. \tag{S17}$$

Defining $\mathcal{C}(t) = e^{-2k^2 V_x(t)}$, we obtain

$$\mathcal{S}_{\text{pos}}(t) = 1 + \frac{4\mathcal{C}(t)}{[1 + \mathcal{C}(t)]^2}. \tag{S18}$$

For the 12$\lambda$-spaced unblockaded superradiant measurement, we use $T = 27$ $\mu$K, $\omega_r = 2\pi \times 70$ kHz, and $t_r = 670$ ns. Following recapture, the position variance oscillates during the 5 $\mu$s cavity detection window. Averaging the predicted scattering rates over this interval gives a superradiant enhancement of approximately $\overline{\mathcal{S}_{\text{pos}}} = 1.5$, compared with the ideal value of 2.

### B. Mixing between magnetic sublevels ($m_F$ mixing)

Superradiant emission requires the atoms to maintain identical optical dipoles and phases throughout the probe pulse. In our experiment, however, the driven $F = 2 \leftrightarrow F' = 3$ transition contains multiple Zeeman sublevels. Spontaneous emission during probing can pump an atom out of the initially prepared $|F = 2,\ m_F = 2\rangle$ state, changing the amplitude, polarization, and phase of its scattered field and thereby reducing the collective enhancement.

We model this effect using two 12-level atoms, including the five $m_F$ states of the $|5S_{1/2},\ F = 2\rangle$ ground manifold and the seven $m'_F$ states of the $|5P_{3/2},\ F' = 3\rangle$ excited manifold. We neglect depumping and subsequent repumping to

reduce the Hilbert-space dimension. These processes introduce additional mixing among $m_F$ states and are therefore expected to further degrade the collective coherence.

In the model, both atoms are positioned at cavity antinodes and coherently coupled to the two transverse cavity polarizations (i.e. two degenerate cavity modes linearly polarized along $z$ and $y$, respectively). The cavity and probe propagation axes are oriented at 45° to the magnetic-field quantization axis. The Hamiltonian includes the Zeeman shifts from the 3 G magnetic field, polarization-dependent probe and cavity couplings with the appropriate Clebsch–Gordan coefficients [S5], and the complete two-atom excitation manifold. We use a cycling-transition probe Rabi frequency $\Omega_p = 2\pi \times 77.6$ MHz and atom–cavity coupling $g = 2\pi \times 2.4$ MHz, with both couplings reduced to account for finite atomic localization at the measured temperature of 27 $\mu$K. We neglect atom motion during probing, as the simulated Zeeman-state populations do not depend strongly on the coupling strengths, and account separately for motional dephasing through the Debye–Waller factor. Free-space spontaneous emission among all allowed Zeeman states and cavity photon loss are included through Lindblad jump operators.

We simulate the 5 $\mu$s probe period using quantum trajectories in QuTiP [S6] for both two atoms and a single atom, averaging over 200 trajectories and integrating the emitted cavity photon number. We define the enhancement $\mathcal{S}_{m_F}$ as the two-atom photon number normalized to twice the corresponding single-atom signal, such that ideal coherent emission gives $\mathcal{S}_{m_F} = 2$. For our experimental parameters, $m_F$-state mixing reduces the enhancement to $\mathcal{S}_{m_F} = 1.7$.

We approximate the reductions in interference contrast from finite atom localization and $m_F$-state mixing as independent. This approximation is supported by the weak dependence of the simulated $m_F$ populations on the probe and cavity coupling strengths. The corresponding interference factors therefore multiply, giving $\mathcal{S}_{\mathrm{tot}} = 1 + (\mathcal{S}_{m_F} - 1)(\mathcal{S}_{\mathrm{pos}} - 1) = 1.4$, which is comparable to our experimentally observed superradiant enhancement.


\end{document}